\documentstyle[preprint,prd,aps,epsf]{revtex}
\begin{document}
\thispagestyle{empty}
\draft
\tighten
\title{\hfill {\small DOE-ER-40757-102} \\
\hfill {\small UTEXAS-HEP-97-16} \\
\hfill {\small MSUHEP-70812} \\
\hfill \\ 
Higgs-photon production at $\mu\bar{\mu}$ colliders}

\author{Ali Abbasabadi$^1$, David Bowser-Chao$^2$, Duane A. Dicus$^3$ and 
Wayne W. Repko$^4$}
\address{$^1$Department of Physical Sciences, Ferris State University, 
Big Rapids, Michigan 49307}
\address{$^2$Department of Physics,
University of Illinois at Chicago, Chicago, Illinois 60607}
\address{$^3$Center for Particle Physics and Department of Physics \\
University of Texas, Austin, Texas 78712}
\address{$^4$Department of Physics and Astronomy \\
Michigan State University, East Lansing, Michigan 48824}

\date{\today}
\maketitle 
\begin{abstract}
We present cross sections for the reaction $\mu\bar{\mu}\rightarrow H\gamma$
over a range of $\mu\bar{\mu}$ collider energies. The amplitudes for this 
process
receive tree level contributions and one-loop contributions, which are of
comparable magnitude. The tree level amplitudes are dominated by helicity
non-flip terms and the one-loop amplitudes are dominated by helicity flip terms.
As a consequence, the interference terms between the tree level and one-loop
contributions are negligible. For a 500 GeV $\mu\bar{\mu}$ collider, the cross
section for $H\gamma$ associated production approaches 0.1 fb.
\end{abstract}

\section{Introduction}
Recently, the possibility of using $\mu\bar{\mu}$ colliders to investigate the
properties of 
Higgs-bosons has received considerable attention \cite{smass,gun}. There are
significant advantages to studying Higgs-bosons with this type of collider,
particularly if the mass is known from its discovery at, say, the LHC or NLC
\cite{gun}. Under these circumstances, the width and branching ratios can be
studied at the Higgs pole.

In the report, we present results for the cross section $\mu\bar{\mu}\rightarrow
H\gamma$ and discuss its use in determining properties of the Standard Model 
Higgs-boson. At a muon collider, unlike an electron-positron collider, the
Higgs-boson-muon coupling is sufficiently large to make the tree level 
amplitude for $H\gamma$ associated production of comparable size to the 
one-loop amplitude \cite{lt}. The contribution of the latter, not included
in Ref.\cite{lt}, is given here together with a discussion of the helicity 
dependence of both the one-loop and tree level amplitudes.

The tree level amplitudes are given in the next section. Section \ref{loop}
contains a summary of the one-loop amplitudes, and this is followed by a
discussion.

\section{Tree level amplitudes}\label{tree}

The tree level diagrams for $\mu\bar{\mu}\rightarrow H\gamma$ are shown in 
Fig. 1, and the resulting amplitude is 
\begin{equation}
{\cal M}^{\rm tree}  = 
\frac{egm_{\mu}}{2m_W}\left[-i\left(\frac{p\!\cdot\!\epsilon^*}{p\!\cdot\!k}
- \frac{\bar{p}\!\cdot\!\epsilon^*}{\bar{p}\!\cdot\!k}\right)\bar{v}(\bar{p})
u(p)\,+\,
\left(\frac{1}{2p\!\cdot\!k} + \frac{1}{2\bar{p}\!\cdot\!k}\right)
\bar{v}(\bar{p})\sigma_{\mu\nu}k_{\nu}u(p)\,\epsilon^*_{\mu}\,\right]\,,
\end{equation}
where $p$ is the momentum of the $\mu$, $\bar{p}$ the momentum of the
$\bar{\mu}$, $k$ the momentum of the $\gamma$ and $\epsilon$ its
polarization. For massless muons, only the helicity non-flip contributions from
the factors $\bar{v}(\bar{p})u(p)$ and $\bar{v}(\bar{p})\sigma_{\mu\nu}u(p)$ are
non-zero. Thus, we expect the helicity flip tree amplitudes to contain a factor
of order $m_{\mu}/E$ relative to the non-flip amplitudes. Explicitly, we
find,
\begin{equation} \label{treeamp}
{\cal M}^{\rm tree}_{\lambda\bar{\lambda}\lambda_{\gamma}} = 
-i\frac{egm_{\mu}}{\sqrt{\displaystyle 2}\,m_W}\left(\frac{1}{2p\!\cdot\!k} + 
\frac{1}{2\bar{p}\!\cdot\!k}\right)\left\{
\begin{array}{lcl}
\sin\theta\left[\lambda_{\gamma}(2|{\bf p}|^2 - E\omega) + |{\bf p}|\omega
\right] &,& \lambda\bar{\lambda} = ++ \\
\sin\theta\left[\lambda_{\gamma}(2|{\bf p}|^2 - E\omega) - |{\bf p}|\omega
\right]&,& \lambda\bar{\lambda} = -- \\
m_{\mu}\,\omega(1 + \lambda_{\gamma}\cos\theta) &,& \lambda\bar{\lambda} = +- 
\\
m_{\mu}\,\omega(1 - \lambda_{\gamma}\cos\theta) &,& \lambda\bar{\lambda} = -+ 
\\
\end{array}
\right.\,,
\end{equation}
where $E$ is the muon energy in the center of mass, $|{\bf p}| = \sqrt{E^2 -
m_{\mu}^2}$, $\omega$ is the photon energy, $\theta$ is the photon scattering
angle and $\lambda_{\gamma} = \pm 1$ is the photon helicity.
It can be seen that the helicity flip amplitudes have a factor of $m_{\mu}$.

\section{One-loop amplitudes}\label{loop}

    The one-loop amplitudes for $\mu\bar{\mu}\rightarrow H\gamma$ receive 
contributions from pole diagrams involving virtual photon and $Z$ exchange and 
from various box diagrams containing muons, gauge bosons and/or Goldstone 
bosons \cite{ab-cdr,ddhr}. 
There are also double pole diagrams whose contribution vanishes. This is 
illustrated in Fig. 2. In the non-linear gauges we chose \cite{ab-cdr}, the 
full amplitude consists of four separately gauge invariant terms: 
a photon pole, a $Z$ pole, $Z$ boxes and $W$ boxes. These amplitudes can be 
written as
\begin{eqnarray}\label{g}
{\cal M}^{\gamma}_{\rm pole} & = &\,\frac{\alpha^2m_W}{\sin\!\theta_W}\,
\bar{v}(\bar{p})\gamma_{\mu}u(p)\left(\frac{
\delta_{\mu\nu}k\!\cdot\!(p + \bar{p}) - k_{\mu}(p + \bar{p})_{\nu}}{s}
\right)\epsilon_{\nu}^*{\cal A}_{\gamma}(s)\,, \\ [4pt]
{\cal M}^{Z}_{\rm pole} & = &\,\frac{\alpha^2m_W}{\sin^3\!\theta_W}\,\bar{v}
(\bar{p})\gamma_{\mu}(v + \gamma_5)u(p)\left(\frac{\delta_{\mu\nu}k\!
\cdot\!(p + \bar{p}) - k_{\mu}(p + \bar{p})_{\nu}}{(s - m_Z^2) + 
im_Z\Gamma_Z}\right)\epsilon_{\nu}^*{\cal A}_{Z}(s)\,, 
\\ [4pt]
{\cal M}_{\rm box}^Z & = &-\frac{\alpha^2m_Z}{4\sin^3\!\theta_W\cos^3\!
\theta_W}\,\bar{v}(\bar{p})\gamma_{\mu}(v + \gamma_5)^2u(p)
\left\{\left[\delta_{\mu\nu}k\!\cdot\!p - k_{\mu}p_{\nu}
\right]{\cal B}_Z(s,t,u)\right.
\nonumber \\
&   &+ \left.\left[\delta_{\mu\nu}k\!\cdot\!\bar{p} - k_{\mu}\bar{p}_{\nu}
\right]{\cal B}_Z(s,u,t)\right\}\epsilon_{\nu}^*\;, \\ [4pt]
\label{w}{\cal M}_{\rm box}^{W} & = &\,\frac{\alpha^2m_W}{2\sin^3\!\theta_W}\,
\bar{v}(\bar{p})\gamma_{\mu}(1 + \gamma_5)^2u(p) 
\left\{\left[\delta_{\mu\nu}k\!\cdot\!p - k_{\mu}p_{\nu}
\right]{\cal B}_W(s,t,u)\right.\nonumber \\
&   &+ \left.\left[\delta_{\mu\nu}k\!\cdot\!\bar{p} - k_{\mu}\bar{p}_{\nu}
\right]{\cal B}_W(s,u,t)\right\}
\epsilon_{\nu}^*\;,
\end{eqnarray}
where $s = -(p + \bar{p})^2,t = -(p - k)^2\,$ and $u = -(\bar{p} - k)^2$. Here,
$v$ denotes the $\mu\bar{\mu}Z$ vector coupling constant, $v = 1 -
4\sin^2\!\theta_W$.
In terms of the scalar functions defined in the appendices of our
previous paper \cite{ab-cdr}, we have \cite{two}
\begin{eqnarray} \label{gam}
{\cal A}_{\gamma}(s) & = & \left\{4(6 + \frac{m_H^2}{m_W^2})
C_{23}(s,m_H^2,m_W^2) - 16C_0(s,m_H^2,m_W^2)\right. \nonumber \\ 
&   &\left.-\,\frac{16}{3}\frac{m_t^2}{m_W^2}\left(4C_{23}(s,m_H^2,m_t^2) 
 -  C_0(s,m_H^2,m_t^2)\right)\right\}\,,\\ [6pt]
\label{z}{\cal A}_Z(s) & = &\; \left\{\left(5 - \tan^2\theta_W
 + \frac{m_H^2}{2m_W^2}(1 - \tan^2\theta_W)\right)C_{23}(s,m_H^2,m_W^2)\right.
 \nonumber \\
&   &\left.+ \left(\tan^2\theta_W - 3\right)C_0(s,m_H^2,m_W^2)\right. \nonumber
\\
&   &\left. - \frac{1}{2}\frac{m^2_t}{m^2_W}
\frac{1 - (8/3)\sin^2\theta_W}{\cos^2\!\theta_W}\left(4C_{23}(s,m_H^2,m_t^2) - 
C_0(s,m_H^2,m_t^2)\right)\right\}\,, \\
{\cal B}_Z(s,t,u) & = & A(s,t,u)\,, \\
{\cal B}_W(s,t,u) & = & A_1(s,t,u) + A_2(s,u,t)\,,
\end{eqnarray}
with $m_t$ denoting the top quark mass. In this case, it is the helicity flip
contributions from the factors $\bar{v}(\bar{p})\gamma_{\mu}u(p)$ and 
$\bar{v}(\bar{p})\gamma_{\mu}\gamma_5 u(p)$ which survive in the
$m_{\mu}\rightarrow 0$ limit. This can be seen by noting that, in the center of
mass, we have
\begin{mathletters}
\label{loopamp}
\begin{eqnarray}
\bar{v}_+(\bar{p})\gamma_{\nu}u_+(p) & = & im_{\mu}(\bar{p} - p)_{\nu}/|{\bf p}|
\,,\\
\bar{v}_+(\bar{p})\gamma_{\nu}\gamma_5 u_+(p) & = & im_{\mu}(\bar{p} + p)_{\nu}
/E \,,\\
\bar{v}_+(\bar{p})\gamma_{\nu}u_-(p) & = & -2\sqrt{2}iE\xi^{(-)}_{\nu} \,,\\
\bar{v}_+(\bar{p})\gamma_{\nu}\gamma_5u_-(p) & = & -2\sqrt{2}i|{\bf p}|
\xi^{(-)}_{\nu} \,,
\end{eqnarray}
\end{mathletters}
with $\xi^{(-)}_{\nu} = (1,-i,0,0)/\sqrt{2}$.
If we define ${\cal M}^{\rm loop}$ as
\begin{equation}
{\cal M}^{\rm loop} = {\cal M}^{\gamma}_{\rm pole} + {\cal M}^{Z}_{\rm pole} + 
{\cal M}^Z_{\rm box} + {\cal M}^W_{\rm box}\,,
\end{equation}
then ${\cal M}^{\rm loop}$ is predominantly helicity flip.

\section{Discussion}

The differential cross section $d\sigma(\mu\bar{\mu}\rightarrow H\gamma)/d\Omega
_{\gamma}$ is given by
\begin{equation}\label{dsig}
\frac{d\sigma(\mu\bar{\mu}\rightarrow H\gamma)}{d\Omega_{\gamma}} =
\frac{1}{256\pi^2}\frac{s - m_H^2}{\beta s^2}\sum_{\rm spin}
|{\cal M}^{\rm tree} + {\cal M}^{\rm loop}|^2\,,
\end{equation}
with $\beta = \sqrt{1 - 4m_{\mu}^2/s}$. When integrating Eq.\,(\ref{dsig}) to
obtain the total cross section, we expect the contributions from
the interference terms to be suppressed, since ${\cal M}^{\rm tree}_{\pm\mp}$ 
and ${\cal M}^{\rm loop}_{\pm\pm}$ both contain an additional factor of 
$m_{\mu}$. This conclusion can only be invalid if the angular integration 
of the muon propagator factors $(1 \pm \beta\cos\theta)^{-1}$ in 
Eqs.\,(\ref{treeamp}) produce inverse powers of $m_{\mu}$. 
This is not the case. For the $++$
or $--$ interference terms, the tree and loop amplitudes contain a factor of
$\sin\theta$, which ensures that the angular integral is well behaved in the
$\beta\rightarrow 1$ limit. The integral of $+-$ and $-+$ interference terms can
produce a factor of $\beta^{-1}$, but this, too, is finite as
$\beta\rightarrow 1$. As a consequence, we can simply add the tree \cite{+-} and
one-loop cross sections to obtain $\sigma(\mu\bar{\mu}\rightarrow H\gamma)$.

The result is illustrated in Fig. 3, where the tree, one-loop and total cross
sections are plotted for several values of $m_H$ as a function of the collider
energy. For collider energies $\sqrt{s} \agt 500$\,GeV, the one-loop
contribution exceeds the tree contribution. Note that Fig. 3 should not be 
taken literally at $\sqrt{s} \approx m_H$, where $\omega\rightarrow 0$ and the 
tree-level process is,in fact, the soft (infrared-divergent) QED correction 
to the resonance process $\mu\bar{\mu} \rightarrow H$. Here, we are concerned 
with production of the Higgs with an observable, relatively hard photon.
In Fig. 4, the total cross
section is shown as a function of $m_H$ for several collider energies. At 500
GeV, luminosities of order 100 fb$^{-1}$ are needed to probe this channel. To
make this statement more precise, we investigated the principal background
$\mu\bar{\mu}\rightarrow b\bar{b}\gamma$ by adapting the amplitudes for
$e\bar{e}\rightarrow \mu\bar{\mu}\gamma$ \cite{eemmg}. In Table I, the 
background
contributions are shown for several cuts on the $b\bar{b}$ invariant mass
$m_{b\bar{b}}$. In addition to these invariant mass cuts, we require the
transverse momenta of the $b$, $\bar{b}$ and $\gamma$ to be greater than 15 GeV,
their rapidities $y$ to be less than 2.5 and the separation $\Delta R$ between
the $\gamma$ and the $b$ and the $\gamma$ and the $\bar{b}$ to be greater than 
0.4. The
background is compared to the signal in Table II for Higgs boson masses of 100
GeV and 200 GeV. This comparison shows that, while not a discovery mode for the
Higgs boson, photon-Higgs associated production can be observed with signal to
square root of background ratios $(S/\sqrt{B})$ greater than 2 when $m_H > 
100$\,GeV at a 500 GeV collider.

Finally, the presence of a large one-loop contribution makes it possible to use
$\mu\bar{\mu}\rightarrow H\gamma$ as a probe of the Higgs-boson coupling
to $W$'s, $Z$'s and top quarks. To determine the sensitivity of the $H\gamma$
cross section to changes in Standard Model couplings, we have varied the
$t\bar{t}H$ coupling by a factor $\lambda$ \cite{hks}. The result is shown in
Fig. 5, where the characteristic feature is the minimum in the cross section
at the Standard Model value $\lambda = 1$. For $\lambda > 1$, the cross section
rises significantly. At a 500 GeV collider, observation of $H\gamma$ production
with a cross section of order 1 fb would indicate some type of anomalous
coupling. 

\acknowledgements

This research was supported in part by the U.S. Department of Energy under
Contract No. DE-FG013-93ER40757 and in part by the National Science Foundation 
under Grant No. PHY-93-07980.

\begin{table}
\begin{center}
\begin{tabular}{cccc}
$\sqrt{s}$ &$45\;{\rm GeV} < m_{b\bar{b}} < \sqrt{s}$&$97.5\;{\rm GeV} < 
m_{b\bar{b}} <102.5\;{\rm GeV}$&$197.5\;{\rm GeV} < m_{b\bar{b}} < 202.5\; 
{\rm GeV}$\\
\hline
500 GeV  & 11.1 fb & $1.16\times 10^{-1}$ fb  & $8.20\times 10^{-2}$ fb \\
1000 GeV & 3.80 fb & $2.69\times 10^{-2}$ fb  & $1.58\times 10^{-2}$ fb \\
2000 GeV & 1.21 fb & $6.83\times 10^{-3}$ fb  & $3.65\times 10^{-3}$ fb \\
4000 GeV & 0.37 fb & $1.81\times 10^{-3}$ fb  & $9.04\times 10^{-4}$ fb
\end{tabular}
\end{center}
\caption{Cross sections for the background process $\mu\bar{\mu}\protect
\rightarrow \gamma b\bar{b}$ are given for several cuts on the $b\bar{b}$
invariant mass $m_{b\bar{b}}$. The last two columns are 5 GeV bins indicating,
respectively, the background associated with a Higgs boson of mass 100 GeV or 
200 GeV.}
\end{table}

\begin{table}
\begin{center}
\begin{tabular}{ccccc}
$\sqrt{s}$ & $\sigma(m_H = 100\;{\rm GeV})$& $S/\sqrt{B}$&$\sigma(m_H = 200\;
{\rm GeV})$& $S/\sqrt{B}$ \\
\hline
500 GeV  &$6.78\times 10^{-2}$\,fb &1.99 & $8.76\times 10^{-2}$\,fb& 3.06 \\
1000 GeV &$2.46\times 10^{-2}$\,fb &1.50 & $3.87\times 10^{-2}$\,fb& 3.08 \\
2000 GeV &$8.76\times 10^{-3}$\,fb &1.06 & $1.04\times 10^{-2}$\,fb& 1.72 \\
4000 GeV &$1.54\times 10^{-3}$\,fb &0.36 & $2.17\times 10^{-3}$\,fb& 0.72
\end{tabular}
\end{center}
\caption{The cross sections for the associated production of 100 GeV and 
200 GeV Higgs bosons are shown together with the ratio of the signal to the 
square root of the background ($S/\protect\sqrt{B}$) for several $\mu\bar{\mu}$ 
collider energies. A luminosity of 100 fb$^{-1}$ is assumed.}
\end{table}

\begin{figure}[h]
\hspace*{1.8in}
\epsfysize=1.3in \epsfbox{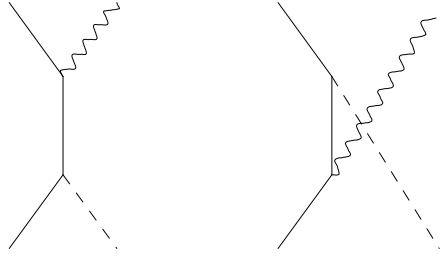}
\vspace{.10in}
\caption{Tree level diagrams for $\mu\bar{\mu}\protect\rightarrow H\gamma$ are
shown.}
\end{figure}

\begin{figure}[h]
\hspace*{1.6in}
\epsfysize=2.0in \epsfbox{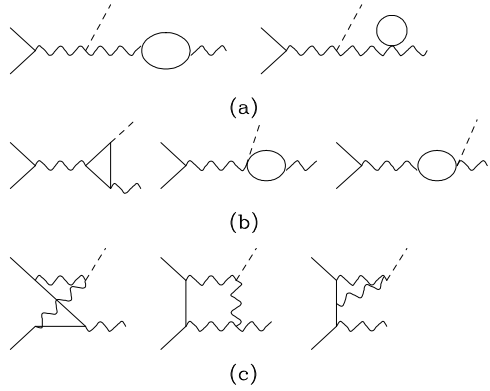}
\vspace{.10in}
\caption{Typical diagrams for the double pole (a), single pole (b) and box (c)
corrections are shown. An external solid line represents a muon, a wavy
line a gauge boson, a dashed line a Higgs boson and an internal solid line a
muon, gauge boson, Goldstone boson or ghost.}
\end{figure}

\begin{figure}[h]
\hspace*{1.2in}
\epsfysize=2.5in \epsfbox{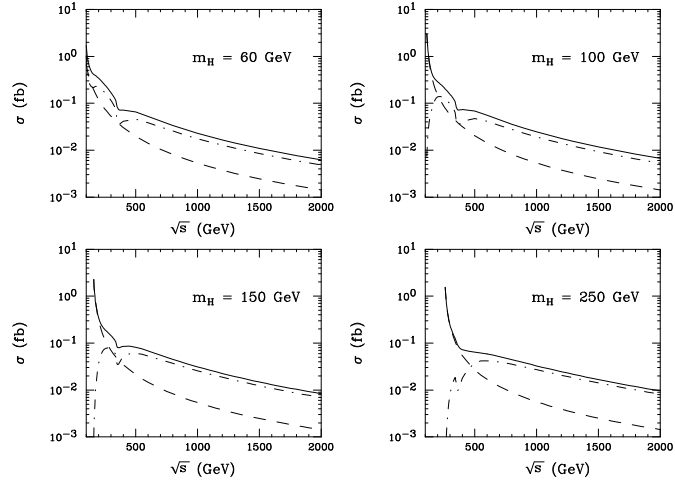}
\vspace{.10in}
\caption{The cross section for $\mu\bar{\mu}\protect\rightarrow H\gamma$
resulting from the sum of the tree level and one-loop amplitudes is given for
several values of $m_H$ by the solid line. In each panel, the dashed line is the
tree level contribution and the dot-dashed line is the one-loop contribution.}
\end{figure}

\begin{figure}[h]
\hspace*{1.2in}
\epsfysize=2.5in \epsfbox{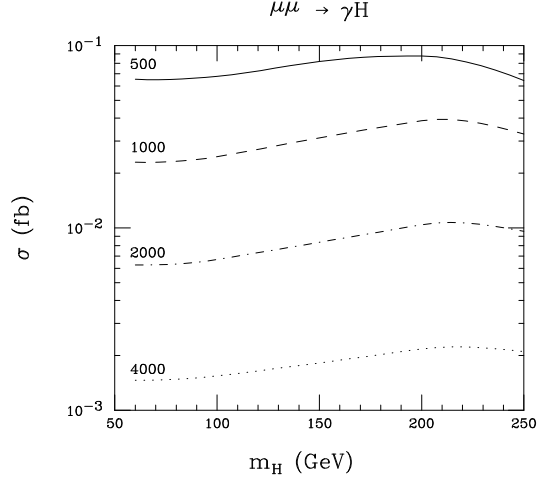}
\vspace{.10in}
\caption{The cross section for $\mu\bar{\mu}\protect\rightarrow H\gamma$
resulting from the sum of the tree level and one-loop amplitudes is given as a
function of $m_H$ for collider energies of 500 GeV, 1000 GeV, 2000 GeV and
4000 GeV.}
\end{figure}

\begin{figure}[h]
\hspace*{1.2in}
\epsfysize=2.5in \epsfbox{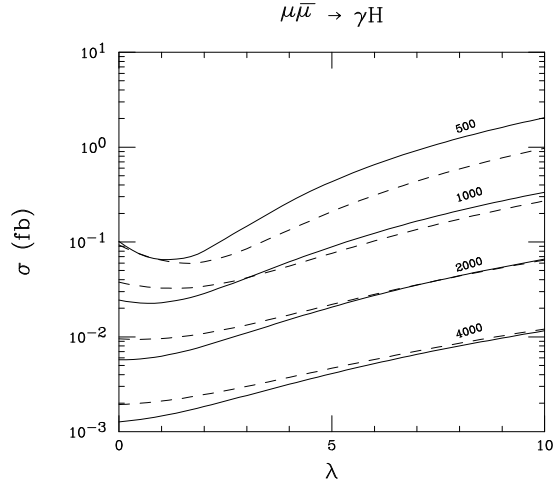}
\vspace{.10in}
\caption{The cross section for $\mu\bar{\mu}\protect\rightarrow H\gamma$
obtained by scaling the Standard Model $t\bar{t}H$ coupling by a factor
$\lambda$ is shown for collider energies of 500 GeV, 1000 GeV, 2000 GeV and 4000
GeV. In each case, the solid line is $m_H = 60$\,GeV and the dashed line is $m_H
= 250$\,GeV.}
\end{figure}

\end{document}